# Low Temperature Structural and Transport Studies of La$_{0.175}$Pr$_{0.45}$Ca$_{0.375}$MnO$_{3-\delta}$

Shivani Sharma, Aga Shahee, Kiran Singh, N. P. Lalla

*UGC-DAE Consortium for scientific research, University Campus, Khandwa Road, Indore-452001,*
*Email:phy.shivanisharma@gmail.com*

**Abstract.** The temperature (T) dependent x-ray diffraction (XRD) and resistivity measurements of La$_{0.175}$Pr$_{0.45}$Ca$_{0.375}$MnO$_{3-\delta}$ (LPCMO) have been performed down to 2K to understand the structural and transport properties. From room temperature down to 220K, LPCMO exists in orthorhombic phase with *Pnma* structure and at T~220K, it transforms to charge ordered (CO) monoclinic phase with *P*2$_1$/*m* structure and remains as it is down to 2K. The CO phase is evident from the occurrence of weak but well defined superlattice peaks in the XRD pattern. This structural transformation is of first order in nature as evident from the phase coexistence across the transition region. These results thus clearly illustrate that LPCMO undergoes a first order structural phase transition from charge disordered orthorhombic phase to CO monoclinic phase at ~220K, consistent with temperature dependent resistivity results. Our structural analysis of T dependent XRD data using Rietveld refinement infers that below 220K, LPCMO forms commensurate CO monoclinic *P*2$_1$/*m* structure with four times structural modulation.

## INTRODUCTION

Manganites possess very interesting properties like CMR effect, multiferroicity etc. and have importance from basic as well as technology point of view. In manganites, La$_{0.175}$Pr$_{0.45}$Ca$_{0.375}$MnO$_{3-\delta}$ (LPCMO) has been studied extensively in the recent years to explore its numerous properties such as metal-insulator transition, charge ordering (CO), phase separation (PS) etc. [1-2]. To study the PS in LPCMO, magnetization measurements have been used extensively [3]. However, the magnetization measurements can distinguish only different magnetic phases like antiferromagnetic and ferro/ferrimagnetic etc. but cannot quantify information about the structural phases involved in PS. The structural studies will provide more insight into the actual origin of PS in any system. For this purpose diffraction techniques like x-ray diffraction (XRD), neutron diffraction and electron microscopy are of direct relevance. LPCMO undergoes well known high temperature (HT)-paramagnetic (PM) to low temperature (LT)-charge ordered (CO) antiferromagnetic (AFM) transition near 220K [3]. As in manganites, strong correlation among different degree of freedom (for example charge, spin and lattice etc.) occurs so it will be interesting to study the structural changes across this magnetic transition. Also in LPCMO, the electronic properties of different phase are quite different so it will be interesting to study its resistivity behaviour during this transition, which will be verified by the LT-XRD results. Here we present a detailed LT-XRD and resistivity studies on LPCMO sample to explore the first order structural transition from HT PM-*Pnma* (orthorhombic) to LT CO-AFM-*P*2$_1$/*m* (monoclinic) phase.

## EXPERIMENTAL DETAILS

The studied sample La$_{0.175}$Pr$_{0.45}$Ca$_{0.375}$MnO$_{3-\delta}$ ($\delta$ = 0.02±0.01) has been prepared via solid state reaction route. The detail procedure of sample preparation has been described elsewhere [4]. Resistivity ($\rho$) measurements have been done during cooling and warming from 150K to 300K with ramp rate of 1K/minute. LT-XRD studies have been done using recently developed low-temperature high magnetic field XRD setup [5]. The XRD data has been recorded at different temperatures during cooling and warming cycle in the angular range of 10° to 110° with scan speed of 2°/minute. Slow scan (0.5°/minute) XRD has also been performed to clearly detect the presence of CO peak.

## RESULTS AND DISCUSSION

The temperature (T) dependent resistivity ($\rho$) behaviour of LPCMO is shown in figure 1. Corresponding to the transition, it shows a kink around ~215K, with a narrow hysteresis in cooling and warming cycles (inset (a) of figure 1), indicating the first order nature of this transition. The transition temperature is determined by the position of sharp kink in the first derivative of $\rho(T)$ (inset (b) of figure 1).

Temperature dependent XRD studies show a structural transition around 220K from *Pnma* to CO monoclinic ($P2_1/m$) phase. The structural changes, emergence of new peaks including CO superlattice peaks is clearly evident in Figure 2 and 3 respectively. Figure 2(a-b) shows a clear first order structural transition at ~220K in cooling and warming cycles. Irreversibility in peaks profiles during cooling and warming cycles clearly indicate phase coexistence and hysteresis across the transition. The presence of CO peak is clearly evident in Figure 3. The CO peaks started appearing below 220K and remained unaltered down to the 2K.

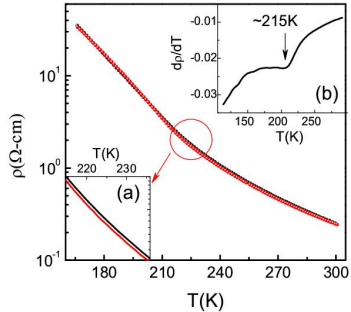

**FIGURE 1.** $\rho$ versus T plot of LPCMO, showing transition at ~215K: inset shows the first derivative of $\rho(T)$ with respect to T.

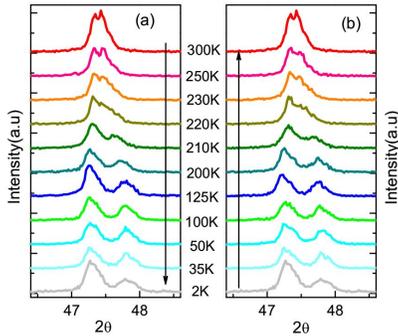

**FIGURE 2.** LT XRD of LPCMO (a) during cooling and (b) warming cycles. Arrow indicated the direction of temperature variation.

Figure 3 represents the Rietveld refined pattern of 80K slow scan XRD data. The inset of figure 3 clearly shows the well fitted CO peak (indicated with arrow).

In manganites, charge and orbital ordering (COO) modulation occurs along [110] direction of the basic perovskite structure. This modulation is a result of transverse displacement of the atomic positions in the structure. We have calculated the modulation (M) from figure 3. The simple vector algebra leads to the M value = $2g^2/(p^2-q^2)$, where g, p and q are the reciprocal lattice vectors directly measured from the XRD pattern converted in reciprocal space ($\text{Å}^{-1}$). The vector g corresponds to the (220) plane of the basic perovskite and p & q correspond to the CO superlattice peaks lying on right and left hand sides of the (220) peak of the basic perovskite. From the XRD data, we have calculated M to be 4 times, which shows that below 220K, *Pnma* phase undergoes commensurate CO transition resulting in $P2_1/m$ structure.

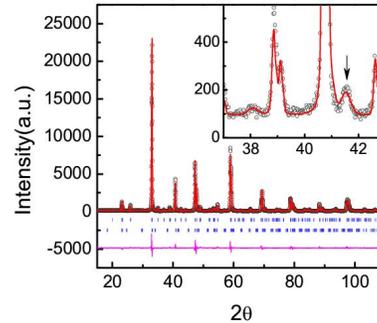

**FIGURE 3.** Rietveld refined pattern of 80K slow scanned (0.5°/minute) XRD data. Inset shows the well fitted CO peak, indicated with black arrow.

## CONCLUSION

Based on the above resistivity and LT-XRD studies we conclude that LPCMO undergoes a first order structural transition from *Pnma* to e CO $P2_1/m$ structure around 220K. The CO was found to be commensurate having 4 time modulation along [110].

## ACKNOWLEDGMENTS

Authors would like to thanks Dr. Rajeev Rawat for resistivity measurements.